\documentclass[nologo,url,11pt,a4paper]{ETHpaper}
\usepackage{graphicx}

\begin{document}
\title{Gender Asymmetries in Reality and Fiction:\\The Bechdel Test of Social Media.}
\titlealternative{Gender Asymmetries in Reality and Fiction: The Bechdel Test of Social Media. \\This is a preprint of a paper
to appear at ICWSM'14 \url{http://www.icwsm.org/2014/index.php}. \\Please cite the conference proceedings when referencing this work.}
\author{David Garcia$^1$, Ingmar Weber$^2$, Venkata Rama Kiran Garimella$^2$}

\address{$^1$ETH Zurich \url{dgarcia@ethz.ch}\\$^2$Qatar Computing Research Institute}
\www{\url{http://www.sg.ethz.ch}}
\maketitle
\centerline{March 25th, 2013}

\begin{abstract}

  The subjective nature of gender inequality motivates the analysis
  and comparison of data from real and fictional human interaction.
  We present a computational extension of the Bechdel test: A popular
  tool to assess if a movie contains a male gender bias, by looking
  for two female characters who discuss about something besides a
  man. We provide the tools to quantify Bechdel scores for both
  genders, and we measure them in movie scripts and large datasets of
  dialogues between users of MySpace and Twitter.  Comparing movies
  and users of social media, we find that movies and Twitter
  conversations have a consistent male bias, which does not appear
  when analyzing MySpace. Furthermore, the narrative of Twitter is
  closer to the movies that do not pass the Bechdel test than to those
  that pass it.

  We link the properties of movies and the users that share trailers
  of those movies. Our analysis reveals some particularities of movies
  that pass the Bechdel test: Their trailers are less popular, female
  users are more likely to share them than male users, and users that
  share them tend to interact less with male users.  Based on our
  datasets, we define gender independence measurements to analyze the
  gender biases of a society, as manifested through digital traces of
  online behavior. Using the profile information of Twitter users, we
  find larger gender independence for urban users in comparison to
  rural ones. Additionally, the asymmetry between genders is larger
  for parents and lower for students. Gender asymmetry varies across
  US states, increasing with higher average income and latitude.  This
  points to the relation between gender inequality and social,
  economical, and cultural factors of a society, and how gender roles
  exist in both fictional narratives and public online dialogues.

\end{abstract}

\section{Introduction}

Gender inequality manifests in objective phenomena, such as salary
differences and inequality in academia \cite{Lariviere2013}, but it
also contains a large subjective component that is not trivial to
measure.  The philosophical works of Simone de Beauvoir describe how
gender inequality is formulated on top of the concept that women are
less rational and more emotional than men \cite{DeBeauvoir1949}. This
points to the subjective and subconscious component of gender
inequality, which prevents many individuals from assessing their own
gender biases in their everyday behavior.  To a great extent, gender
inequality is exercised but not consciously reflected upon, creating a
pattern of biases that everyone experiences but nobody names
\cite{Fredan1963}.  Gender bias, as part of the culture and ideology
of a society, manifest in subconscious behavior and in the fictions
created and consumed by that society \cite{Zizek1989}.  The emerging
field of \emph{culturomics} aims at a quantitative understanding of
culture at a large scale, first being applied to cultural trends in
literature \cite{Michel2011}, but also applicable to other traces of
human culture such as voting in song contests \cite{Garcia2013},
search trends \cite{Preis2012}, and movies \cite{Danescu2011}.

In 1985, Alison Bechdel published the comic strip named ``The Rule'',
in which a female character formulates her rules to be interested in
watching a movie: \emph{It has to contain at least two women in it,
  who talk to each other, about something besides a man}
\cite{Bechdel1985}. While intended as a punchline about gender roles
in commercial movies, it served as an inspiration to critically think
about the role of women in fiction.  Such formulation of a test for
gender biases became popularly know as \emph{the Bechdel
  test}\footnote{\url{en.wikipedia.org/wiki/Bechdel_test}}, and is
usually applied to analyze tropes in mass
media\footnote{\url{www.youtube.com/watch?v=bLF6sAAMb4s}}.  Whether a
movie passes the test is nothing more than an anecdote, but the
systematic analysis of a set of movies can reveal the gender bias of
the movie industry.  The Bechdel test is used by Swedish cinemas as a
rating to highlight a male bias, in a similar manner as it is done
with violence and nudity.  Previous research using this test showed
gender bias when teaching social studies \cite{Scheiner-Fisher2012},
and motivated the application of computational approaches to analyze
gender roles in fiction \cite{Lawrence2011}.

The current volume of online communication creates a record of human
interaction very similar to a massive movie, in which millions of
individuals leave digital traces in the same way as the characters of
a movie talk in a script. This resemblance between reality and fiction
is the base of the theory of behavioral scripts \cite{Bower1979}, used
to analyze subconscious biases through patterns of social interaction
\cite{Shapiro2010}. For example, linguistic coordination appears in
both movie scripts \cite{Danescu2011}, and Twitter dialogues
\cite{Danescu2011a}.  Furthermore, bidirectional interaction in
Twitter has been proved useful for computational social science,
testing theories about the assortativity of subjective well-being
\cite{Bollen2011}, about cognitive constraints like Dunbar's number
\cite{Goncalves2011}, and about conventions and social influence in
Twitter \cite{Kooti2012}. Gender roles in emotional expression appear
in MySpace dialogues \cite{Thelwall2009}, and gender-aligned
linguistic structures have been found in Facebook status updates
\cite{Schwartz2013}.  One of our aims is to compare the patterns of
dependence across genders in movies and online dialogues, assessing
the question of whether our everyday life, as pictured in our online
interaction, would be close to passing the Bechdel test or not.

Large datasets offer the chance to analyze human behavior at a scale
and granularity difficult to achieve in experimental or survey
studies.  At the scale of whole societies, digital traces allow the
analysis of cultural features, such as future orientation through
search trends \cite{Preis2012}, and the similarity between cultures
through song contest votes \cite{Garcia2013}.  At the level of
individual behavior, online datasets allowed the measure of intrinsic
biases, such as that we tend to use more positive than negative words
\cite{Garcia2012}, that we have a tendency to share information with
strong emotional content \cite{Pfitzner2012}, and that apparently
irrelevant actions, such as Facebook likes, reveal relevant patterns
of our personality \cite{Kosinski2013}.

We introduce a quantitative extension of the Bechdel test, to measure
female and male independence in the script of a movie and the digital
dialogues of a population. We calculate these metrics based on the
amounts of dialogues between individuals of the same gender that do
not contain references to the other gender.  In our approach, we keep
a symmetric analysis of male and female users and characters,
quantifying asymmetries without any presumed point of view.  We
combine these metrics with information about geographic location,
personal profile, and movies viewed by groups Twitter users, to take
one step further in understanding the conditions under which gender
biases appear.  We test the role of climate in the gender stereotypes
of a culture, as suggested by previous works that found a relation
between the future orientation and the geographic location of a
culture \cite{Zimbardo2008}. Similarly as the ability to plan ahead is
encouraged by adverse climate, it can also encourage gender biases in
which males behave more independently and less emotionally attached.
We test this theory, known as the \emph{disposable male}
\cite{Farrell1996}, through the hypothesis that male independence
increases with distance to the equator.  Additionally, we explore the
relation between economic factors and female independence, measuring
the relation between the average income of states of the US and the
gender independence of its female Twitter users.

The centralized nature of movies plays a key role in the persistence
of gender inequality in a society, which is part of the concept of
cultural hegemony \cite{Lears1985}.  This case is particularly
important in movies aimed at children, which are known to have a
gender bias that disempowers female characters \cite{Pollitt1991}.  On
the other hand, the lack of central control in the content of online
media offers the chance of gender unbiased interaction, as conjectured
by cyberfeminist theories \cite{Hawthorne1999,Fox2006}. In this
article, we quantify the presence of these biases in dialogues from
movies and Twitter, testing if these patterns prevail in online
communication or are only present in mass media.

\section{Analytical Setup}

To assess the questions mentioned above, we require three main
datasources: references from social media to movies, movie script
information, and dialogues involving the users that shared information
about the movies. Furthermore, we need a set of tools to process these
datasets in order to identify the gender of users and actors, to
detect gender references from social media messages and movie lines,
and to group them into dialogues over which we can analyze gender
asymmetries. In the following, we outline our datasources and the
methods we used for our analysis.

\subsection{Movie scripts, trailers and Bechdel test data}

From an initial dataset of YouTube videos \cite{Abisheva2014}, we
extracted a set of 16,142 videos with titles that contain the word
``trailer'' or ``teaser'', among the categories of Movies and
Entertainment. Removing trailer related terms, we matched these videos
to titles of movies in the Open Movie
Database\footnote{\url{www.omdbapi.com/}} (OMDB), selecting the pairs
in which the string similarity between the titles was above
80\%. After a manual inspection of the results, our dataset contained
704 trailers for 493 movies in 2,970 Twitter shares. In addition, we
had the amount of views, likes, and dislikes for these trailers, as
retrieved in June, 2013.

We downloaded scripts from the Internet Movie Script
Database\footnote{\url{www.imsdb.com/}} as text files with the content
of each movie, similarly to previous works \cite{Danescu2011a}.  To
disambiguate the title of each movie on the site, we first automated a
search through OMDB, matching the title of each movie with the IMDB
identifier of the first result, filtering when title string similarity
was above 80\% and the year of release differed in no more than one
year. Once each movie script was in text form, we identified character
lines and scene cuts using the standard syntax used to process
screenwriting markup languages, as used by screenwriting software like
Fountain\footnote{\url{fountain.io/syntax}}. We constructed dialogues
between characters following the sequence of their lines, and using
the scene cuts as additional explicit separators between dialogues.
Second, we developed a set of python scripts to download the Bechdel
test results form bechdeltest.com, in a similar manner as previously
done for visualization purposes\footnote{\url{http://bit.ly/1c3QF1g}}.
We gathered Bechdel test data for 470 movies with an unique IMDB
identifier, processing the information in bechdeltest.com to extract
two values: a test result $b$ indicating how many rules of the test
are passed by a movie (from 0 to 3), and an amount of disagreements in
the comments of the users of the site.

\subsection{MySpace and Twitter dialogue retrieval}

Using the service provided by Gnip, we collected a set of public
tweets from the period between June 1st and 6th, 2013. Using the list
of YouTube videos mentioned above, we found a set of 536,835 users
that shared at least one of the trailers in our dataset. From among
those users, we extracted their location information, which we matched
with Yahoo!\ Placemaker\footnote{\url{developer.yahoo.com/boss/geo/}}
to select those in the US. As a result, we got 69,606 \emph{ego} users
in the US that shared at least one trailer.  For each one of these
users, we retrieved their history of tweets with a limitation of 3,200
tweets per user, and their lists of followers and followees with the
limitation of 5,000 users in each list.  From this set of friendships,
we identified those that interacted with the users in our database if
they exchanged at least 10 mentions. Filtering by US location
again, we extended this set of users with an additional 107,645
\emph{alter} users, for which we also got up to 3,200 of their most
recent tweets. This composes a track of the recent history of
interactions between more than 170,000 users, over which we
constructed dialogues based on the more than 300 million tweets they
wrote. We complement this Twitter dataset with a set of MySpace
messages provided in previous research \cite{Thelwall2009}, covering
conversations of users in the US and the UK.

We constructed a network of dialogues between Twitter users, based on
the mentioning functionality of @-tags. For each tweet, we know its
text, the user that created it, the time of creation, and the users
that were mentioned in it. The full track of tweets between two users
composes a set of dialogues, in which the users talked about different
topics at various moments in time. There is no explicit sign in the
tweets between two users that indicates the beginning and the end of a
dialogue, which motivates the application of heuristics
\cite{Danescu2011a} and semantic-based machine learning methods
\cite{Ritter2010}.

\begin{figure}[b]
  \centering
  \includegraphics[width=0.95\textwidth]{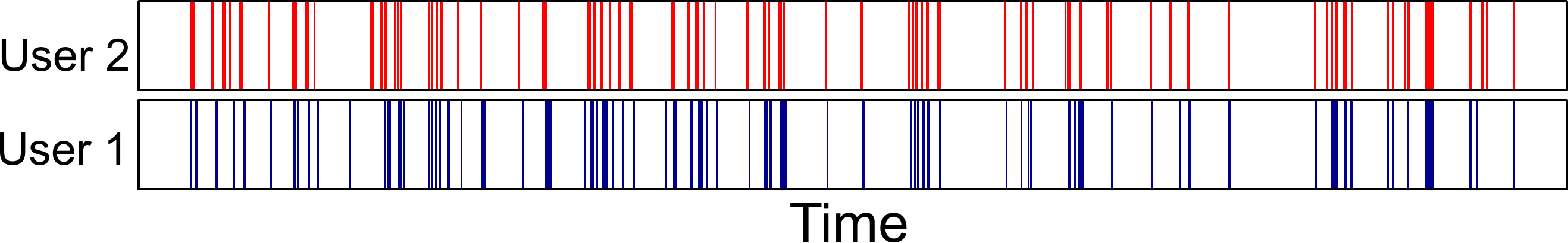} \caption{Example
    of timestamps of the tweets written by two users mentioning each
    other in a period of two weeks.\label{fig:DiscussionExample}}
\end{figure}

For our case, we use the theory of bimodality in human communication
\cite{Wu2010}, which states that there are two modes in the
interaction between pairs of people: an intra-burst mode in which
messages follow each other in a dialogue after very short periods, and
an inter-burst mode of long silences between dialogues. Figure
\ref{fig:DiscussionExample} shows an example of the timestamps of the
tweets exchanged between a pair of Twitter users.  The head of the
distribution of times between messages of a pair of users follows a
power-law close to $P(t) \sim t^{-3/2}$ for time intervals in
$[t_{min}, \tau]$, and the tail is closer to an exponential
distribution of the form $P(t)\sim e^{-\beta t}$ for $t>\tau$.  Note
that this bimodal definition is different than a power-law with
exponential cutoff, which is typically used to control for finite-size
effects \cite{Clauset2009,Wu2010}.  The cutoff time $\tau$ gives us an
estimate of the time scale of correlations inside a dialogue.  We
estimated its value in MySpace and Twitter through a modified version
of the maximum likelihood technique for power-laws \cite{Clauset2009},
correcting for the fact that the fit is to the head and not the tail
of the distribution. This method minimizes the Kolmogorov-Smirnov
criterion of distribution equality giving power-law exponents of
$1.48$ in Twitter and $1.43$ in MySpace, very close to previous
empirical studies of this kind of distributions in IRC channels
\cite{Garas2012}. This method allowed us to compute estimates of the
cutoff value $\tau$ between both distributions, which are of 9.1 hours
in Twitter and 7.7 hours in MySpace.

\begin{figure}[t]
  \centerline{\includegraphics[width=0.7\textwidth]{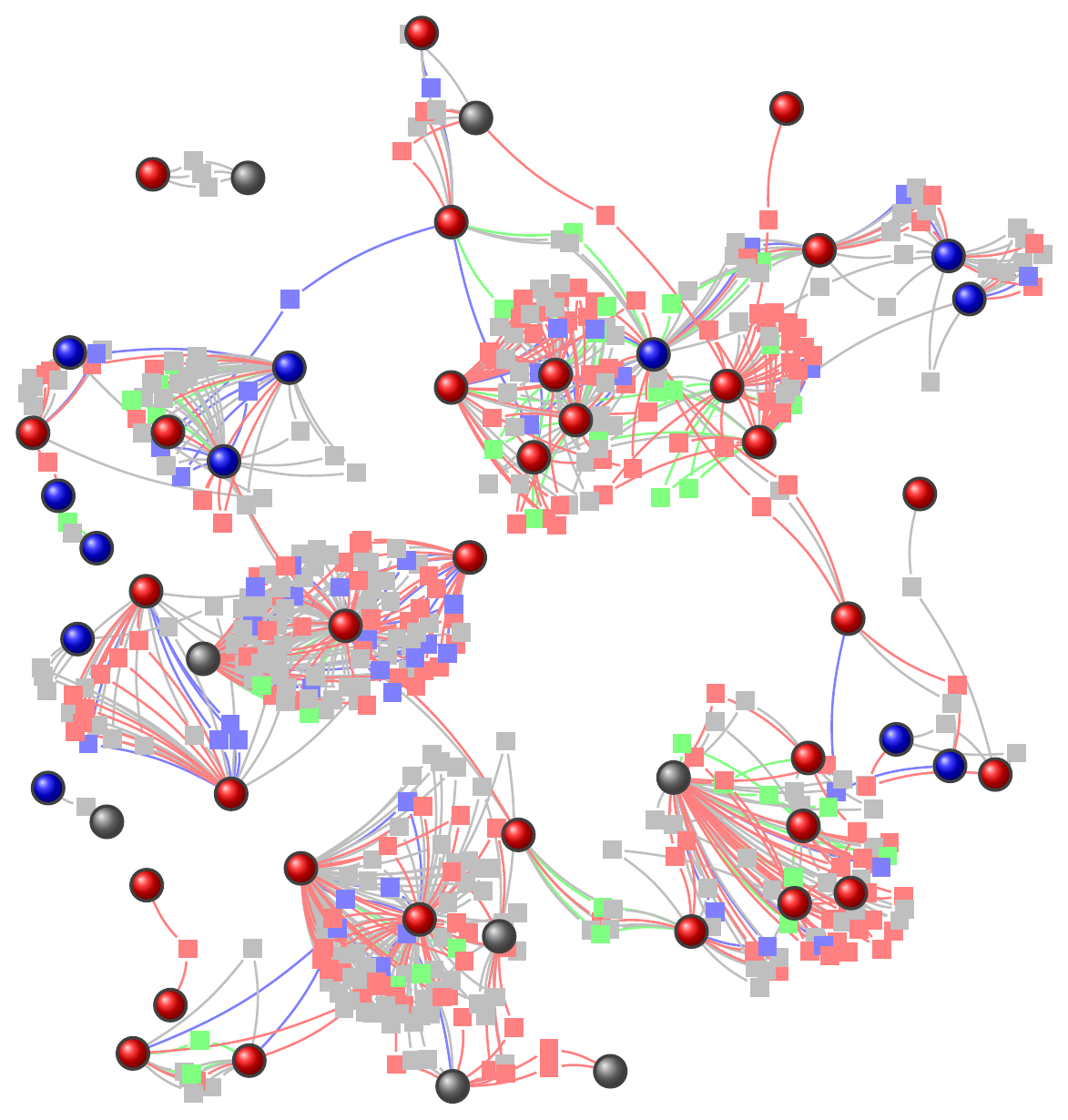}}
  \caption{Networks of dialogues between users located in the city of
    Ann Arbor, MI.  Circular nodes represent users, colored blue if
    they are female, red if male, and gray if their gender could not
    be determined. Squares represent dialogues between users, colored
    red if the dialogue contains male references, blue if it contains
    female references, green if it refers to both male and female, and
    gray if there is no reference to any gender in the dialogue. In
    this subnetwork $B_F$= 0.06 and
    $B_M$=0.36. \label{fig:CityNetwork}}
\end{figure}

We used the time cutoffs to separate dialogues, applying the rule that
times of silence longer that the cutoff indicate the end of a
dialogue. As a result, our Twitter dataset is composed of 2,240,787
dialogues and the MySpace dataset of 3,263 dialogues.  An illustration
of a subnetwork of the resulting data is Figure~\ref{fig:CityNetwork},
where we show the dialogues between Twitter users that declared to
live in Ann Arbor, Michigan.

\subsection{Profile information processing}

We retrieved demographic information from the Twitter profile of each
user, looking for keywords that signal personal information, as
done in previous research \cite{Mislove2011,Pennacchiotti2011}.  This
way, we identify the users as likely father, mother, or student if they
include terms related to parenthood or studying in their
profile. Their location information allowed us to find their city and
state within the US, which we matched versus the list of the 100
largest cities in the US
\footnote{\url{www.city-data.com/top1.html}} to identify urban
and rural users. We identified the gender of a user through first name
matching against the history of names in the
US\footnote{\url{www.ssa.gov/oact/babynames/limits.html}},
classifying the gender of a user in the same way as previously done
for Twitter \cite{Mislove2011} and for authors of research papers
\cite{Lariviere2013}.

For each movie, we gathered the cast list from IMDB, and then we
looked for the terms ``actor'' and ``actress'' in the IMDB profile
of the actors playing each character, which determines their gender
\cite{Danescu2011a}. This data is not only useful to determine the
genders of the characters in our movie dataset, but also serve as a
ground truth to evaluate the dictionary technique we used for Twitter
users. In total, we found 4,970 actors and 2,486 actresses, with 154
unknown (writers and directors).  We applied the above gender
detection technique, which gave us the unknown gender class 33\% of
the time. Note that these actors and actresses do not need to live in
the US, and that can have artistic names like \emph{``Snopp Dogg''},
who appears in our dataset. Precision values are 0.894 for detecting
males and 0.844 for detecting females, and recall values are 0.582 for
males and 0.595 for females. The similarity in these values indicates
that this tool does not introduce a bias that changes the ratio of
male and female users. Nevertheless, improvements are possible, not
only in the lexica, but also introducing machine learning techniques
that find more complicated naming patterns \cite{Pennacchiotti2011}.

\subsection{Detection of gender references}

We modify the above gender detection technique to find which dialogues
include male and female references. First, we modified the gender
lexica used above, filtering out English dictionary words, like
``Faith'', and toponyms. Second, we disambiguated the gender of names
that can appear in both genders by the frequency of appearance in each
gender in the US. If a name is used for a gender at least than 5 times
more often than for the other gender, we assign it to the lexicon of
the gender with the highest frequency, or we remove it otherwise.
Second, we add common feminine words like ``her'', and masculine words
like ``him''. For each dialogue composed by a set of tweets, this
technique will classify as containing references to males, females,
both, or none depending on the presence of common words and names
associated each gender.

In our study, we apply this technique for a very particular subset of
dialogues, aiming only at the decision whether male-male dialogues
contain female references, and whether female-female dialogues contain
male references. To provide an initial validation beyond intuition, we
set up a small experiment to estimate the quality of the method. We
extracted a random set of 100 dialogues between male Twitter users
that were classified as containing female references and 100 as not
containing them, and a random set of 100 dialogues between female
users that were classified as containing female references and 100 as
not containing them. Manual annotation using third party evaluations
gave us an approximation of the ground truth, under the impossibility
of surveying the actual users involved in the dialogues.

\begin{table}[t]
\centering
\begin{tabular}{|c|c|c|c|c|c|} \hline
  Problem  & Pre. (1)  & Rec. (1) & Pre. (0)  & Rec. (0)\\
  \hline
  f in D(M,M) &     0.69  &    0.986   &     0.99  &    0.76 \\\hline 
  m in D(F,F) &    0.8 &     0.964    &    0.97 &     0.83    \\\hline
\end{tabular}
\caption{Validation of detection of the presence (1) or absence (0) of gender references in dialogues.\label{tab:GenderReferenceRes}}
\end{table}

Table~\ref{tab:GenderReferenceRes} shows the results of this
validation.  Recall values for finding cross-gender references in
male-male and female-female dialogues are very high: i.e.\ the
dictionary method tends not to miss many true gender
references. Precision value for finding female references is lower due
to misunderstanding surnames as first names, or outliers in the
meaning of words, like ``Marylin Manson''.  The higher precision of
male reference detection implies that it is more likely that the tool
makes a mistake when tagging a male-male dialogue as not having female
references than its female counterpart. This means that the tool
introduces certain bias that lowers the amount of detected dialogues
between males that do not contain female references, which indicates
that our findings of a generalized larger male independence are robust
to the precision of our tools.

\section{The Bechdel Score}

\begin{figure}[ht]
\centerline{\includegraphics[width=0.7\textwidth]{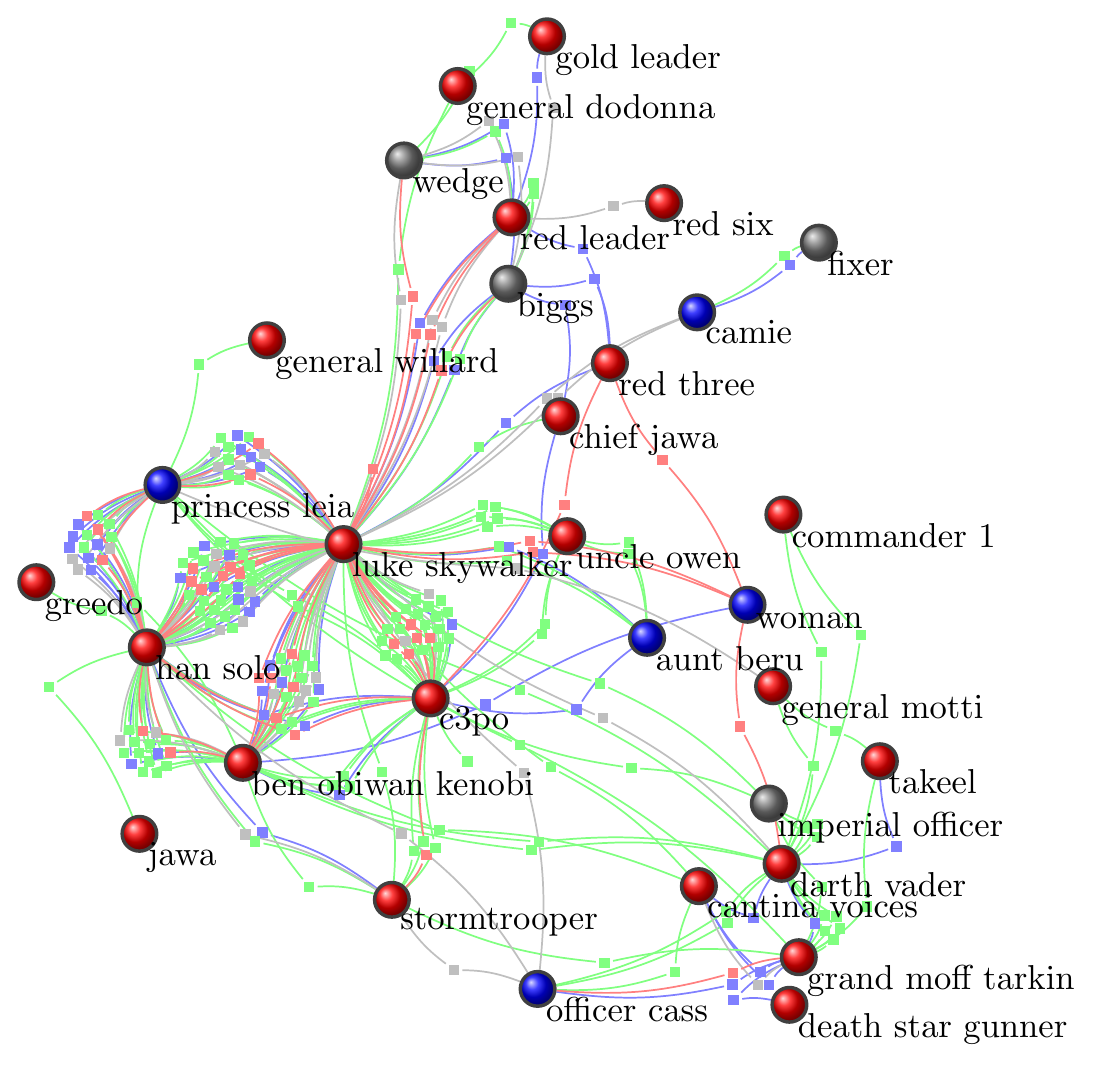}}
\caption{Character network of \emph{Star Wars Episode IV: A New Hope}.
  Colors and nodes represent characters, dialogues, and the genders
  involved as in Figure \ref{fig:CityNetwork}.  For this movie, $B_F=
  0$, $B_M= 0.163$. \label{fig:MovieNetwork}}
\end{figure}

After processing the movie scripts, each movie corresponds to a
network in which characters are connected through
dialogues. Figure~\ref{fig:MovieNetwork} shows as an example the
network extracted from the script of the first \emph{Star Wars} movie.
For each movie, we have a set of dialogues $D$, where each dialogue is
a tuple $\{g_1,g_2,m,f\}$, in which $g_1$ and $g_2$ are the genders of
the characters in the dialogue ($M$ for male, $F$ for female, and $U$
for unknown), and $m$ and $f$ are binary values set to 1 if the
dialogue includes a male reference or a female reference respectively,
and 0 otherwise.  We define the set of dialogues of a movie with
certain values in the tuple as $D(g_1,g_2,m,f)$. For example, all the
dialogues between pairs of female characters that do not contain
references to any gender are $D(F,F,0,0)$.

To quantify the level of independence of one gender from the other, we
define a score metric that can be computed for each gender.  We
calculate the female Bechdel score $B_F$ of a movie as the ratio of
dialogues between female characters that do not contain male
references, over the whole set of dialogues of a movie, and its male
counterpart $B_M$:
\begin{equation}
  B_F = \frac{|D(F,F,0,*)|}{|D|} \qquad B_M = \frac{|D(M,M,*,0)|}{|D|}, 
  \label{eq:bechdelscore}
\end{equation}
where the $*$ in the equation represents any possible value for the
variable.  We quantify the extent of the asymmetry between genders
displayed in a movie through the comparison of $B_F$ and $B_M$. In the
example of Figure~\ref{fig:MovieNetwork}, many male characters
dialogue with each other, and some of these interactions do not
contain female references (red or gray). Our method detected the
presence of 5 female characters, and the absence of any dialogue in
which two of these female characters talk to each other, giving a
female Bechdel score of 0.

\begin{figure}[t]
  \centering
  \includegraphics[width=0.95\textwidth]{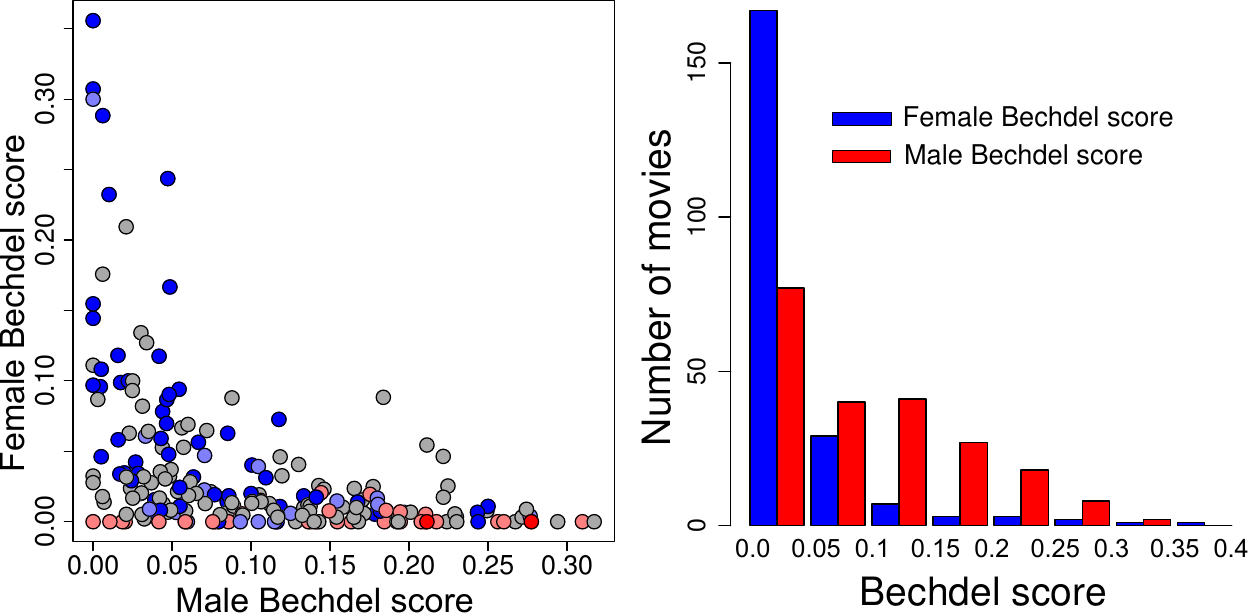}
  \caption{ Left: Scatter plot of the male and female Bechdel scores
    of each movie. Dark blue dots are movies with $b=3$, light blue
    with $b=2$, light red with $b=1$, dark red with $b=0$, and gray if
    no reliable Bechdel test information was available.  Right:
    distribution of $B_M$ and $B_F$.\label{fig:BechdelCartesian}}
\end{figure}

We computed the values of $B_F$ and $B_M$ for the set of 213 movies in
our dataset, as shown in the left panel of Figure
\ref{fig:BechdelCartesian}. Point colors correspond to the results of
the Bechdel test in \url{bechdeltest.com}, coloring gray the movies
that do not appear on the website, or that have disagreements in the
user comments. Dark blue dots are movies that pass the Bechdel test
($b=3$), which populate the area above $B_F=0$. The line where $B_F=0$
or very low is populated with movies that do not pass the Bechdel
test, some of them which do not even have two female characters
($b=0$).

Both measures, the Bechdel test and the Bechdel score, are subject to
certain degree of uncertainty.  The precise value of the Bechdel score
is subject to a certain heuristic component introduced by the gender
identification technique, as explained in the analytical setup
section.  On the other hand, the manual annotation of the Bechdel test
often has disagreements, as the test contains a set of implicit
assumptions about which characters should be included and what
constitutes a dialogue.  Nevertheless, there is a clear relation
between our computation of Bechdel scores and the manual annotations
of the Bechdel test: Wilcoxon tests show that movies that pass the
test have higher $B_F$ by 0.026, ($p<10^{-9}$) and lower $B_M$ by
0.051 ($p<10^{-3}$).

The density of points in the area with low $B_F$ reveals that there is
a generalized bias towards less independence of female characters.
This effect becomes clear in the distributions of $B_F$ and $B_M$, as
shown in the right panel of Figure \ref{fig:BechdelCartesian}.  Male
Bechdel scores are higher than female Bechdel scores, with a
difference between medians of 0.07 ($p<10^{-15}$). This confirms the
observation that movies in English tend to portray female characters
as more dependent on male characters than vice versa.

The videos of the trailers for these 213 movies provide information
about their popularity in terms of their amount of views, likes, and
dislikes. We tested the equality of the distribution of these
popularity values for the set of movies that pass and do not pass the
Bechdel test. We found that movies that do not pass the test have more
views in their trailers, with a difference between medians of about
265,000 views ($p=0.0015$). They also receive more likes, as the
difference between the median amount of likes is 329
($p=0.012$), but the hypothesis that they contain the same amount of dislikes could
not be rejected ($p=0.061$). This points to the fact that movies with
a male gender bias are more popular in YouTube than those that show a
more symmetric view of gender interdependence.

\begin{figure}[b]
  \centering
  \includegraphics[width=0.6\textwidth]{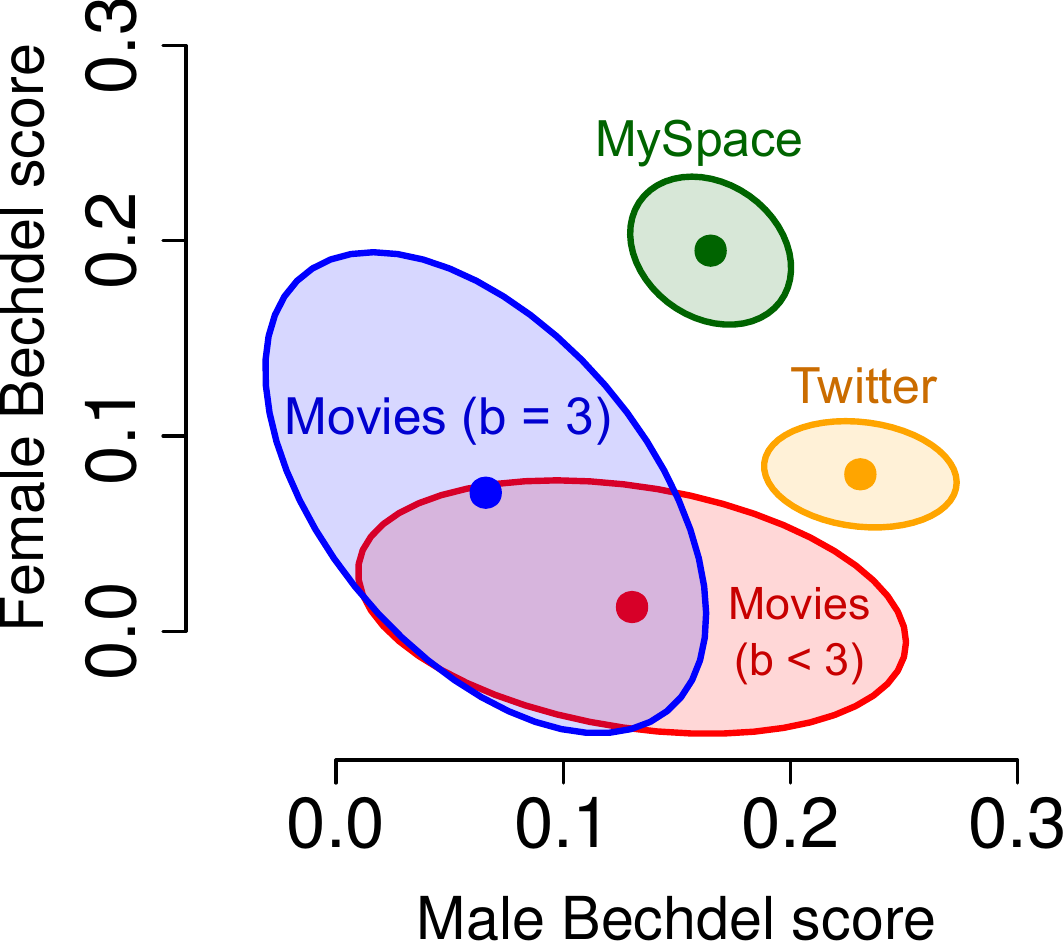}
  \caption{Yellow and green dots show centroids of $B_F$ and $B_M$ for
    subsets of dialogues in MySpace and Twitter of the same size as
    the average movie. Blue and red dots show the centroids for movies
    that pass the Bechdel test (b=3) and that do not pass it
    (b$<$3). Ellipses show the distance of one standard deviation
    around bivariate means.\label{fig:Ellipses}}
\end{figure}

\section{Gender Roles in Movies and Social Media}

We computed Bechdel scores for the MySpace and Twitter datasets,
creating a bootstrap partition of the set of dialogues into subsets of
the same size as the average movie (225). For each of these samples,
we computed $B_F$ and $B_M$, estimating this way the size and
uncertainty in the gender bias present in each online medium.

 Figure~\ref{fig:Ellipses} shows a comparison
with the same procedure applied to movies that pass ($b=3$) and do not
pass the Bechdel test ($b<3$), computing the centroid and standard
deviation in the space of both Bechdel scores.  An initial observation
reveals that Twitter has a strong bias towards male independence,
being closer to movies that do not pass the test.  To verify this, we
computed Wilcoxon distances between the Bechdel scores in MySpace,
Twitter, and the two sets of movies, reported in
Table~\ref{tab:BechdelDistances}.  Weighting both Bechdel scores
equally, Twitter has a smaller Euclidean distance to movies that do
not pass the test (0.127) than those that pass it (0.18).

One of the reasons why Twitter has such large male Bechdel score is
the uneven ratio of users of each gender. In line with previous
research \cite{Mislove2011}, 64\% of the users in our dataset are
male.  In order to use social media data to analyze societies at
large, we need to account the unequal volumes of activity and presence
of each gender. For this reason, we decompose the Bechdel score in two
metrics: dialogue imbalance to measure the difference in activity
between both genders, and gender independence to estimate the real
fraction of dialogues that do not contain references to the other
gender in a larger population.

We define female dialogue imbalance $X_F$ as the likelihood that one
endpoint of a discussion is a male user, given that the other endpoint
is female. The male extension of this metric is the male dialogue
imbalance $X_M$, computed as the ratio of male-male dialogues over all
dialogues that involve males:
\begin{eqnarray}
    X_F = \frac{ |D(F,M,*,*) \cup  D(M,F,*,*)|} {
      |D(F,*,*,*)\cup D(*,F,*,*)|} \nonumber \\
    X_M = \frac{ |D(M,M,*,*)|} {
      |D(M,*,*,*)\cup D(*,M,*,*)|} 
    \label{eq:imbalance}
\end{eqnarray}

Assuming that Twitter gives us a sample of conversations with unequal
gender ratios, but taken from a society with equal amount of male and
female individuals, we can rescale our computation to Bechdel scores
into gender independence ratios. We define female gender independence
$I_F$ and male gender independence $I_M$ as the ratio of dialogues that
do not include male references from among female-female dialogues, and
the ratio of dialogues that do not include female references from
among male-male dialogues:
\begin{equation}
    I_F = \frac{|D(F,F,0,*)|}{|D(F,F,*,*)|} \qquad I_M = \frac{|D(M,M,*,0)|}{|D(M,M,*,*)|} 
    \label{eq:independence}
\end{equation}
Through these metrics, given a sufficiently large sample of dialogues,
we can estimate the degree of independence of one gender with respect
to the other, as articulated in their dialogues in social media.

\begin{table}[t]
\centering
\begin{tabular}{|c|c|c|c|} \hline
 Twitter  & $\Delta B_F$  & $\Delta B_M$ &  Euclidean Distance  \\
 \hline
$b=3$ &       0.035   &       0.179  &     0.18   \\\hline
$b<3$ &         0.075 &       0.103 &              0.127    \\\hline\hline
 MySpace   & $\Delta B_F$  & $\Delta B_M$ &  Euclidean Distance  \\ \hline
$b=3$  &       0.146   &       0.112  &    0.184  \\ \hline
$b<3$ &         0.18  &         -   &                 0.18  \\ \hline
\end{tabular}
\caption{Distances in Bechdel scores between social network datasets and movies
  that pass and do not pass the Bechdel Test. 
  All differences are significant under  $p<10^{-5}$, with the
  exception of $B_M$ between MySpace and movies (b$<$3) for which
  there was no significant difference $p=0.09$.\label{tab:BechdelDistances}}
\end{table}
\begin{figure}[b]
  \centering
  \includegraphics[width=0.95\textwidth]{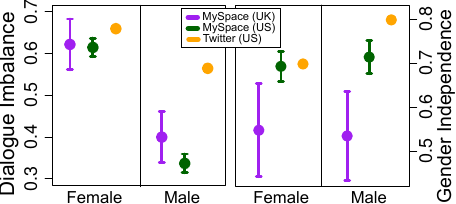}
  \caption{Dialogue imbalance and gender independence ratios for
    Twitter and MySpace UK and US. Error bars show 95\% confidence
    intervals , and the size of the error bars for Twitter is smaller
    than point size. \label{fig:GendIndep}}
\end{figure}
The MySpace dataset was constructed in a balanced way, downloading
dialogues from similar amounts of male and female users. The left
panel of Figure~\ref{fig:GendIndep} shows that the value of $X_F$
above 0.5 and of $X_M$ below 0.5 reveals a pattern of disassortativity
in which users of different genders tend to interact, revealing the
use of MySpace for dating at the time of the crawl
\cite{Thelwall2009}.  The high ratio of male users in Twitter is
captured by the dialogue imbalance metric, which shows the large
likelihood for a dialogue to involve a male.  For this reason, both
$X_M$ and $X_F$ are above 0.5, indicating that the large male Bechdel
score of Twitter is due to this size difference.

The right panel of Figure~\ref{fig:GendIndep} shows the male and
female gender independence ratios, for both MySpace datasets (UK and
US) and Twitter. While in MySpace these independence ratios were
similar for males and females, the male independence in Twitter is
significantly larger than its female counterpart. This shows that the
asymmetry between the independence of male and female users of Twitter
remains, even when we control for the amount of gender-aligned
dialogues. This implies a clear bias towards male independence in
Twitter, despite of not being centrally controlled by any agent.  This
is not the case for MySpace, where tests of equality of proportions
did not allow us to conclude any differences in $I_F$ and $I_M$. This
points to the limited size of the MySpace dataset, which is not a
limitation for our Twitter dialogues dataset. We require large
datasets to measure $I_F$ and $I_M$, which is the reason why they are
not applicable to individual movies or small samples from social
media.

\section{Relating Asymmetry in Movies and Twitter}

After name matching between YouTube Trailers and movies, our Twitter
dataset contains 1,741 trailer shares made my male users and 588 by
female users.  Out of those movies, we have Bechdel test information
for 662 shares from male users and 294 shares from female users.  For
a subset of those movies we also have script information and Bechdel
scores, accounting for 264 shares from males and 86 shares from
females.  From those users, we compute $X_M$ and $X_F$ in their ego
network, if they participate in at least 25 dialogues. In the
following, we show the relation between the gender biases present in a
movie, measured through Bechdel test results and Bechdel scores, and
the ratios of each gender of the users and their dialogue imbalance.

The female Bechdel score of the movies in shares from female users are
in general larger than those shared by male users. A Wilcoxon test
comparing both distributions rejects the hypothesis that they are the
same ($p=0.011$). The distance between medians is $0.0047$, which is a
relevant size in the $B_F$ scale, indicating that shares from female
users are about movies with $B_F$ 45.5\% higher than $B_F$ in shares
by male users.  The opposite pattern with male Bechdel scores was not
significant ($0.081$) to conclude that male or female Twitter users
shared movies with higher $B_M$.

\begin{figure}[t]
  \centerline{
  \includegraphics[width=0.47\textwidth]{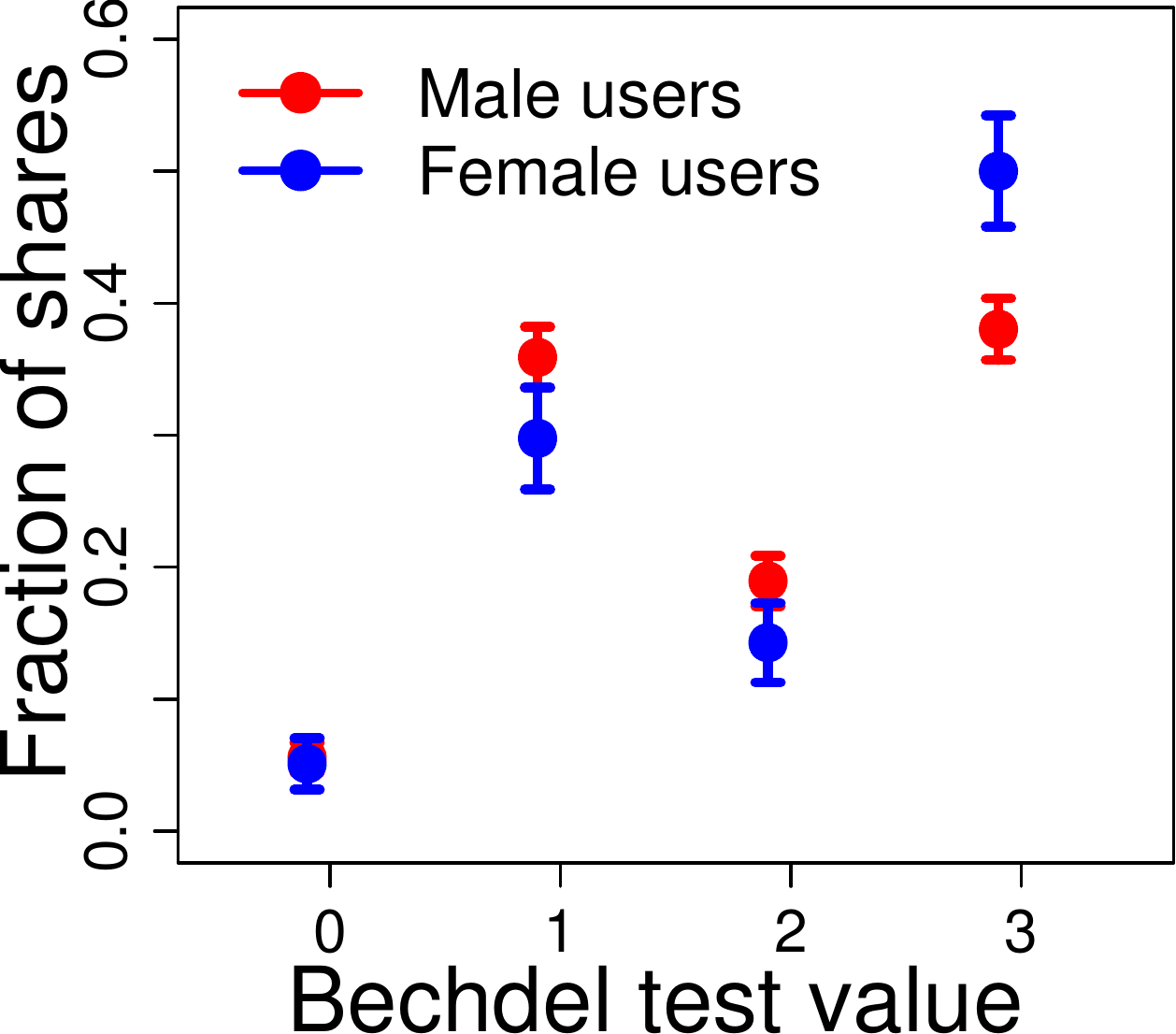} \hfill
  \includegraphics[width=0.47\textwidth]{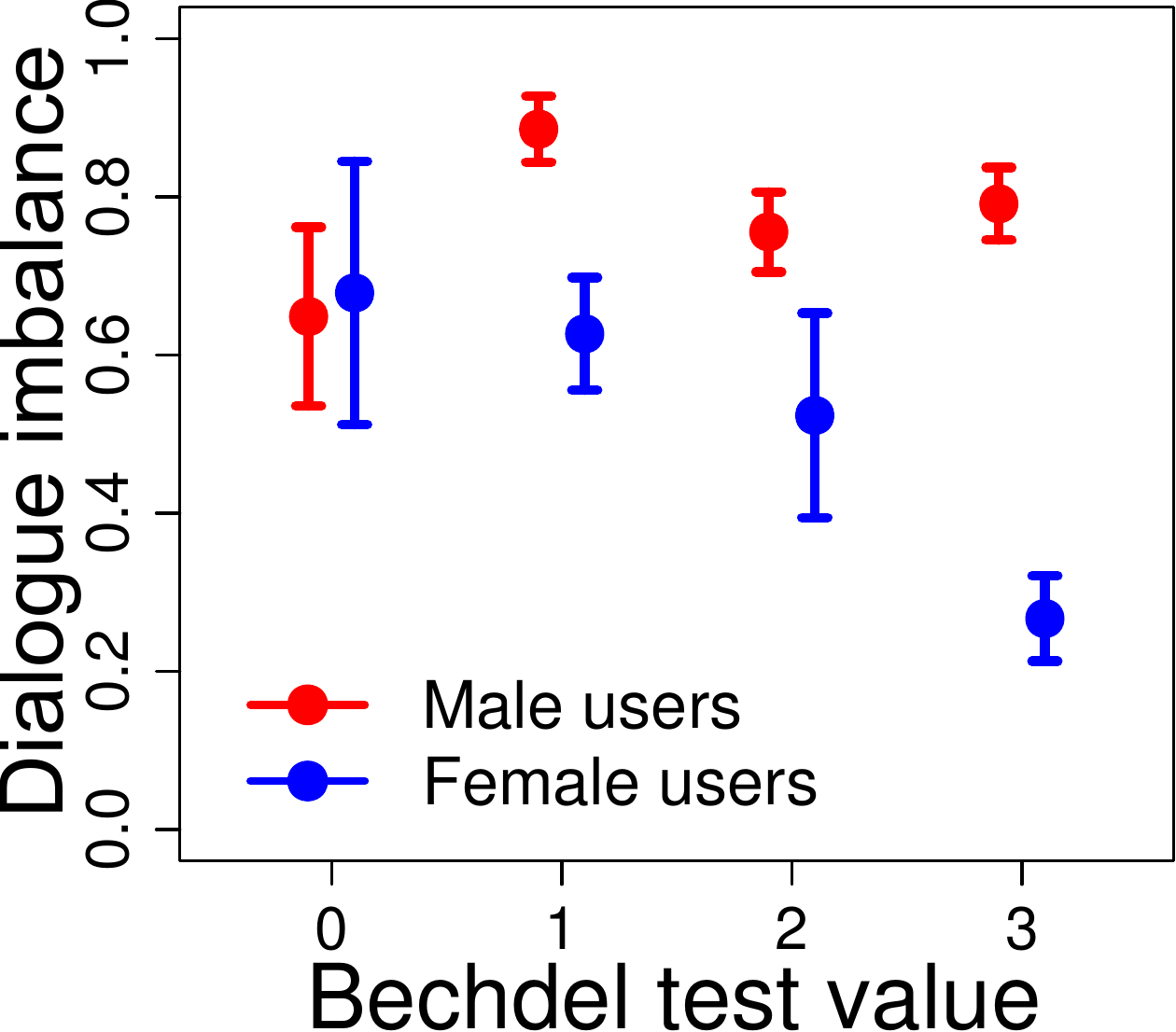}}
\caption{ The left panel shows the ratios of each Bechdel test value
  for the movies shared my male and by female Twitter users. The right
  panel shows the median discussion imbalance of users that share
  movies of each value of the Bechdel test. \label{fig:TwitterMovies}}
\end{figure}

The numeric value of the Bechdel test ($b$) of movies in shares from
female users had a larger value than those on the shares by male users
($p<10^{-4}$), revealing that women share movies that pass more rules
of the Bechdel test.  Furthermore, a share from a female user is more
likely to be about a movie that passes all three rules of the Bechdel
test.  We computed the ratio of shares about movies that pass the test
from all shares from women, and the same among the shares from men.  A
$\chi^2$ test on these ratios indicates that women are more likely to
share movies that pass the Bechdel test ($p<10^{-6}$), by an increase
of 0.12, 30\% more than the likelihood for male users. This relation
can be seen in the left panel of Figure~\ref{fig:TwitterMovies}, where
we show the fraction of shares about movies with certain Bechdel test
value, over all the shares from male and female users. The most
striking difference is at 3, pointing to the higher chances that
female users share movies that pass the test.

We computed the dialogue imbalance of the users sharing each movie
trailer, and compared them across genders and across movies that pass
or not the test.  The dialogue imbalance of female users that share
movies that pass the Bechdel test is lower than for female users
sharing movies that do not pass the test ($p=0.013$).  This was not
the case for male users ($p=0.111$), showing no relation between
Bechdel test results and the behavior of male Twitter users.  Across
genders, the difference between $X_M$ and $X_F$ for the users that
share movies that pass the Bechdel test is significant ($p<10^{-6}$),
and of magnitude 0.42 on the total dialogue imbalance scale. This
difference is not significant among males and females that share
movies that do not pass the test ($p=0.4265$), indicating that there
is a shift away from interaction with men in the population of women
that share movies that pass the test. The right panel of
Figure~\ref{fig:TwitterMovies} shows this effect, where $X_F$ and
$X_M$ diverge when computed over the dialogues in users that shared
movies that pass the Bechdel test. This analysis shows that there are
relations between the behavior of female Twitter users and the movies
they consume and share, but male users do not show any variation with
respect to the portrayal of female roles in these movies.

\section{Profile Factors in Gender Independence}

The information displayed in Twitter profiles allowed us to extract
two variables of the personal life of a user: if they mentioned in
their profile that they are mothers or fathers and whether they are
students. We use this information to analyze the gender independence
present in dialogues between users of a particular kind, in comparison
with the rest of dialogues present in Twitter.

\begin{figure}[t]
  \centerline{
  \includegraphics[width=0.9\textwidth]{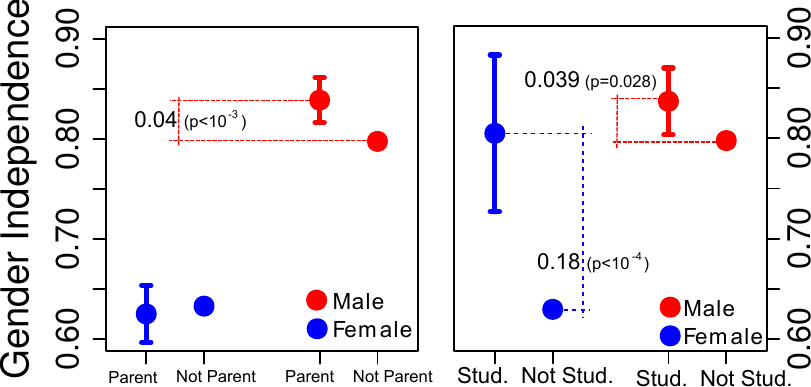}}
\caption{ Gender independence ratios for the set of dialogues between
  parents and non-parents (left) and between students and
  non-students. Error bars show 95\% confidence intervals, possibly
  smaller than point size.\label{fig:ParentsStudents}}
\end{figure}

There is no significant difference in $I_F$ when comparing dialogues
between mothers and the rest of female-female dialogues.  On the other
hand, fathers showed a significantly higher male gender independence,
as shown in the left panel of Figure \ref{fig:ParentsStudents}. This
indicates that publicly articulated dialogues between fathers tend to
mention women less often.  The gender independence in the discussions
between pairs of two male students and two female students have higher
values of $I_F$ and $I_M$, when compared to dialogues between users
not identified as students.  The right panel of Figure
\ref{fig:ParentsStudents} shows that the difference between Female
students and the rest is very large, to the point of not having any
significant difference with $I_M$ between male students.  The gender
asymmetries we find in the Twitter population are not evident among
students, suggesting that gender roles are less prominent within that
subset of the population.

\section{Geographical factors}

\begin{figure*}[t]
  \centering
  \includegraphics[width=0.97\textwidth]{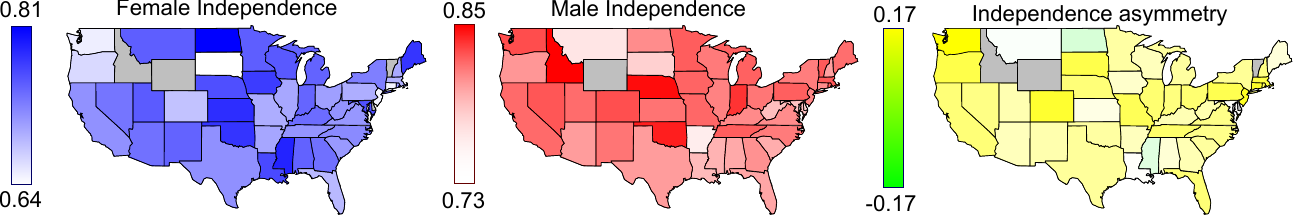} \hfill
  \caption{State maps of gender independence: Each state is colored
    according to $I_F$ (left), $I_M$ (middle) and $I_M-I_F$
    (right). Gray states contained less than 50 gender aligned
    dialogues to have a reliable estimation of $I_F$ or
    $I_M$. \label{fig:StateMaps}}
\end{figure*}

The location information of each Twitter user allows us to build the
networks for all the users located in each state of the United
States. An example of this kind of filtering is
Figure~\ref{fig:CityNetwork}, where only the users located in Ann
Arbor are displayed. This way, for each state we have a set of
dialogues, which we use to measure male and female gender
independence.  Figure~\ref{fig:StateMaps} show $I_F$ and $I_M$ of each
state in the continuous United States, as well as gender asymmetry
computed as $I_M-I_F$. It is noticeable that most of the states have
positive asymmetry, with the exception of Hawaii, Mississippi,
Montana, and North Dakota.

\begin{figure}[b]
  \centering
  \includegraphics[width=0.47\textwidth]{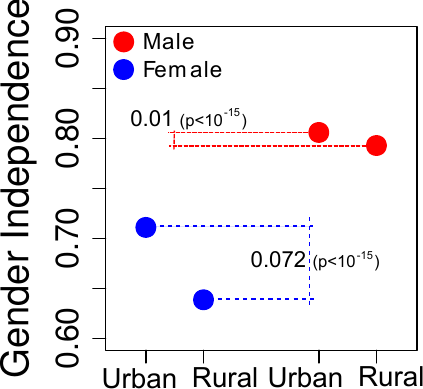}
  \caption{Gender Independence ratios of urban and rural users. Error
    bars show 95\% confidence intervals, and are smaller than point
    size. \label{fig:Urban}}
\end{figure}

Given the location of each user, we tag them as urban if they live in
one of the largest 100 cities of the United States, and rural if they
live in a smaller city, as explained in the analytical setup
section. Figure~\ref{fig:Urban} shows $I_F$ and $I_M$ for discussions between urban
users and between rural users. Both male and female urban users show
larger gender independence, but this difference is stronger between
urban and rural females. Nevertheless, $I_F$ is still significantly
lower than $I_M$ in the urban Twitter population.

\begin{figure}[t]
  \centerline{
  \includegraphics[width=0.9\textwidth]{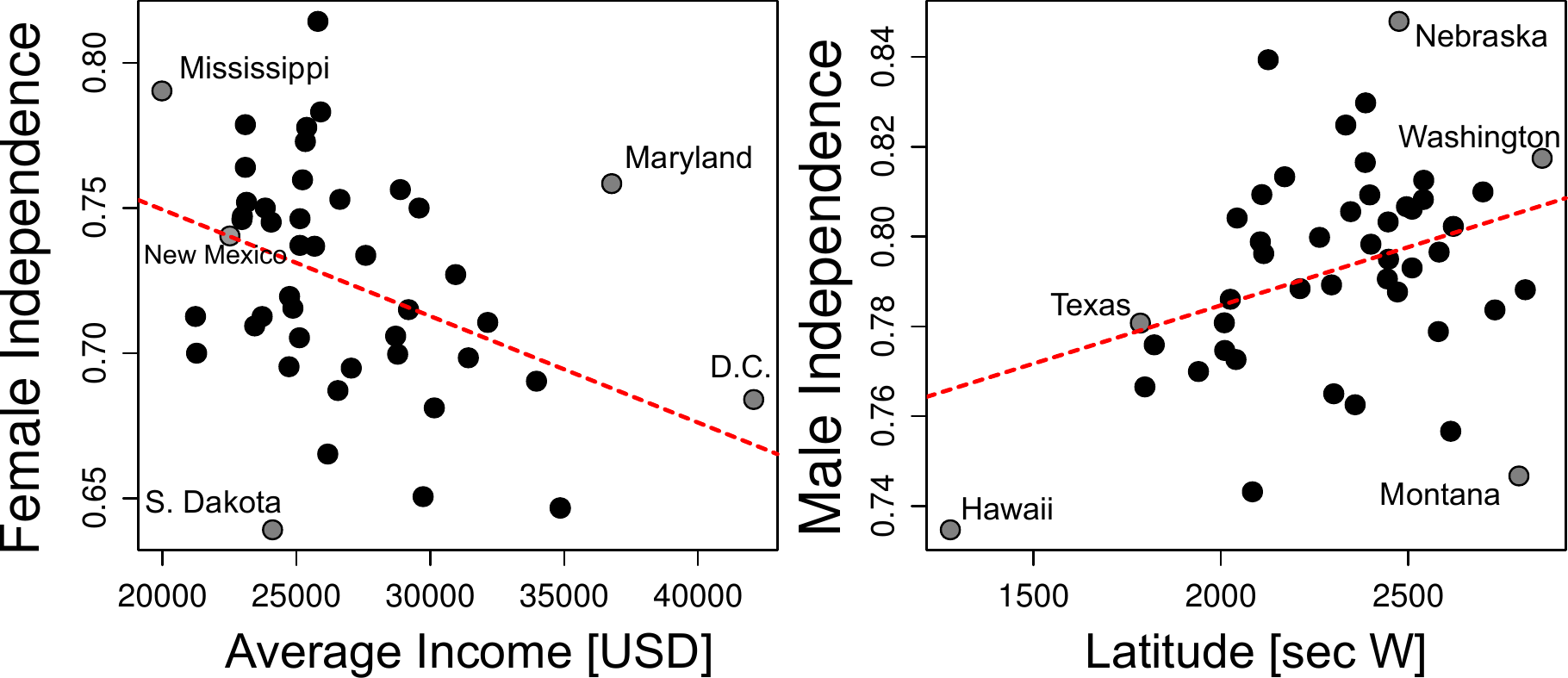}}
\caption{The left panel shows the scatter plot of $I_F$ versus state
  average income. The right panel shows the scatter plot of $I_M$ and
  latitude of the largest city of the state. Red dashed lines show
  linear regression trends.  \label{fig:AsymmetryDeps}}
\end{figure}

Finally, we investigate the relation between other economic and
geographic factors with both $I_F$ and $I_B$. From the national
census, we gathered average income of each state, and the Gini index
of income inequality. For each state, we also located the latitude and
longitude of the largest city, measured in seconds West from the
Greenwich meridian and North from the Equator, in order to test the
role of economy and climate in female and male independence.

$I_F$ has a correlation coefficient with average income of $-0.4$ and
$p=0.0054$, stronger than with any other metric including male
independence. To evaluate if this result is a confound with any other
metric, we calculated partial Pearson correlation coefficients
$\rho(I_F, X|Y)$ where $X$ is average income and $Y$ is each of the
other variables ($I_M$, Gini index, latitude and longitude). All
partial correlations were negative and significant, being the least
significant when controlling for Gini index ($-0.32$ $p=0.028$). This
shows that the dialogues between females in US states with higher
income are more likely to include male references. The left panel of
Figure \ref{fig:AsymmetryDeps} shows the scatter plot of these values.
The outliers of Figure \ref{fig:AsymmetryDeps} call for a closer
analysis of the statistical properties and robustness of our results.
A Shapiro-Wilk test of normality on $I_F$ does not allow us to reject
the null hypothesis that it is normally distributed ($p=0.67$), but
the opposite is true for average income ($p=0.00125$). For this
reason, we replicate the above correlation analysis by computing
Spearman's correlation coefficient, which tests for a monotonous
relation, by reducing the leverage of outliers through a rank
transformation of the data. Spearman's correlation between $I_F$ and
average income was significant and negative ($\rho_s=-0.37$,
$p=0.0107$). 

Male independence is correlated with latitude, with a correlation
coefficient of $0.34$ ($p=0.02$). The same partial correlations
analysis as with $I_F$ reveals positive and significant correlations
when controlling for all the other factors. The least significant of
these is when controlling for average income ($0.31$, $p=0.03$),
showing that latitude is the most related factor with male
independence.  As shown in the right panel of Figure~%
\ref{fig:AsymmetryDeps}, states that are farther from the equator have
male dialogues that are less likely to contain female references.
Normality tests for $I_M$ and latitude did not provide evidence that
they are not normally distributed, as Shapiro-Wilk tests did not
reject the null hypothesis for both $I_M$ ($p=0.87$) and latitude
($p=0.11$). Nevertheless, we tested the statistical robustness of our
finding by computing Spearman's correlation between $I_M$ and
latitude, finding a significant positive correlation of $0.29$
($p=0.04436$).

\section{Discussion}

We presented a study that combines data from movie scripts, trailers,
and casts, and Twitter and MySpace users, including their profile
information and the dialogues among them. We designed a set of metrics
to measure gender biases in the sets of dialogues in movies and social
media, to explore the relations between gender roles in fiction and
reality. Starting from an equal approach to male and female
independence in movies, we verified the existence of a generalized
bias in which female characters are shown as dependent on male
characters. Furthermore, the trailers of male biased movies are more
popular, and the movies shared by Twitter users are related to their
profiles and patterns of interaction.  While we did not find a
difference for male users, female users are more likely to share
movies with high female Bechdel scores, and to interact less with with
male users if they share movies that pass the Bechdel test.  This
indicates that female Twitter users are attracted to movies in which
women are shown less dependent on men, but also that the audiences
might be starting to be aware of the results of the Bechdel test
itself.

We compare the gender biases in Twitter and MySpace with our metrics
for movies, finding that Twitter contains a male bias not only in
amount of users, but also in a lower degree of female gender
independence.  This points to the possibility that some design
decisions of Twitter might create undesired effects
\cite{Goodman2012}, such as hindering female users to engage in the
community in the same way as males do. In addition, the biases present
in public dialogues in Twitter are not radically different from those
in movies.  The decentralized nature of Twitter has not led to a
gender unbiased interaction with respect to mass media, and the
asymmetric pattern of lower female gender independence is also present
in everyday online public interaction.  This similarity between
reality and fiction can be explained by two mechanisms: i) the gender
roles present in fiction, including movies, influence our behavior and
gender bias, or ii) movies reflect patterns of gender dependence
present in real life.  It is also possible that there is a combination
of both mechanisms, in which a feedback loop makes movies reflect
certain gender bias in everyday life, but also perpetuate gender
inequality through the influence of movies in human culture. In any
case, such subconscious biases are a component of ideology and 
contribute to the creation of inequality at a large scale
\cite{Shapiro2010}, in the same way as very small racial preferences
can lead to segregation \cite{Schelling1971}. We find that certain
personal factors are related to the gender independence of Twitter
users. This is particularly strong as we did not find evidence for a
presence of gender bias among students, and we found urban users to be
more gender independent that rural ones, especially for women.

We calculated gender independence values across states in the US,
finding a generalized pattern of asymmetry towards lower female
independence. We found a significant correlation between male
independence and latitude that is consistent with the theory of the
disposable male, which predicates that males behave more independently
due to the presence of adverse conditions, including climate.
However, this result is constrained to US states, and further work on
a wider range of societies is necessary to understand the relation
between climate and gender asymmetries.  In addition, we found a
negative relation between female gender independence and average
income, counter intuitively to the concept that female emancipation
increases the workforce and the productivity of a society.  Note that
our finding is structural, not measuring changes in average income or
GDP due to changes in female independence, and that they are
consistent with the observation that richer countries show larger
gender bias in scientific production \cite{Lariviere2013}. One
possible explanation lies in the difference between the political
discourse regarding gender independence and the subjective behavior of
a society: labeling oneself as liberal in gender policies does not
necessarily imply an absence of a gender bias in everyday behavior.
On the other hand, this result also points to the role of Twitter in
society, as publicly articulated dialogues might emphasize certain
ideals of gender equality in the places where they are needed the
most.  All these questions are empirically testable in future
research, in particular if focused on individual income and education
levels, world views, and gender biases quantified in observed
behavior.

\section{Acknowledgements}
The authors would like to acknowledge Gnip for providing data on
Twitter activity. DG was funded by the Swiss National Science
Foundation (CR21I1\_1464991).

\end{document}